\documentclass[aps,prl,10pt,twocolumn,showpacs,superscriptaddress,longbibliography]{revtex4-2}
\usepackage{amssymb,graphicx,color,amsmath}
\usepackage[utf8x]{inputenc}
\usepackage{amsmath}
\definecolor{lightblue}{rgb}{0.17,0.39,1}
\usepackage[bookmarks, colorlinks=true, breaklinks]{hyperref}
\usepackage[normalem]{ulem}
\hypersetup{linkcolor=lightblue, citecolor=lightblue, filecolor=black, urlcolor=lightblue}

\newcommand{\redtext}[1]{}
\newcommand{\bluetext}{} 
							
\begin{document}

\title{Precursor to Quantum Criticality in Ce-Au-Al Quasicrystal Approximants}

\author{A.~Khansili}
\email[Email: ]{akash.khansili@fysik.su.se}
\affiliation{Department of Physics, Stockholm University, SE-106\/91 Stockholm, Sweden}

\author{Y.-C.~Huang}
\affiliation{Department of Chemistry, Uppsala University, SE-751\/21 Uppsala, Sweden}

\author{U.~Häussermann}
\affiliation{Department of Materials and Environmental Chemistry, Stockholm University, SE-106\/91 Stockholm, Sweden}

\author{C.~Pay~Gomez}
\affiliation{Department of Chemistry, Uppsala University, SE-751\/21 Uppsala, Sweden}

\author{A.~Rydh}
\email[Email: ]{andreas.rydh@fysik.su.se}
\affiliation{Department of Physics, Stockholm University, SE-106\/91 Stockholm, Sweden}

\begin{abstract}
Rare-earth element containing aperiodic quasicrystals and their related periodic approximant crystals can exhibit non-trivial physical properties at low temperatures. Here, we investigate the 1/1 and 2/1 approximant crystal phases of the Ce-Au-Al system by studying the ac-susceptibility and specific heat at low temperatures and in magnetic fields up to 12\,T. We find that these systems display signs of quantum criticality similar to the observations in other claimed quantum critical systems, including the related Yb-Au-Al quasicrystal. In particular, the ac-susceptibility at low temperatures shows a diverging behavior $\chi \propto 1/T$ as the temperature decreases as well as cutoff-behavior in magnetic field. {\bluetext{Notably, the field dependence of $\chi$ closely resembles that of quantum critical systems.}} However, the ac-susceptibility both in zero and nonzero magnetic fields can be understood from the {\redtext{Zeeman}} splitting of a ground state Kramers doublet of Ce$^{3+}$.
The high-temperature Curie-Weiss fit yields an effective magnetic moment of approximately 2.54$\mu_{\mathrm{B}}$ per Ce for both approximant systems, which is reduced to $\sim$2.0$\mu_{\mathrm{B}}$ at temperatures below 10\,K. {\redtext{Notably, this field dependence closely resembles that of quantum critical systems.}}
The low-temperature specific heat is dominated by {\redtext{this}} {\bluetext{the}} Schottky anomaly originating from {\redtext{a}} {\bluetext{the}} splitting of the Ce$^{3+}$ {\redtext{ground state}}  Kramers doublet{\bluetext{, resulting in an entropy of $R\ln 2$ at around 10\,K}}.

\end{abstract}

\date{\today}
\maketitle

\section{Introduction}
Understanding the physical properties of aperiodic quasicrystals (QCs) is challenging, since QCs cannot be easily treated with conventional methods \cite{Levine1984, Levine1986, Goldman1991, Tsai1999}. The structurally and compositionally closely related approximant crystals (ACs), however, are built of the same atomic clusters as QCs but possess periodicity in 3D. Experimentally observed are 1/1 and higher order 2/1 ACs. Especially in the latter, the spacial arrangement of cluster building units connects closely to QCs \cite{Takakura2007}. An important question is whether QCs possess specific physical properties that are exclusive to their long-range quasiperiod order \cite{Wessel2003, Goldman2013, Goldman2014, Thiem2015, Takeuchi2023}. In this respect, quantum critical behavior has been reported in the icosahedral QC Yb-Au-Al (Yb$_{15}$Au$_{51}$Al$_{34}$) \cite{deguchi2012quantum, Watanuki2012}, which shows a divergence of several physical properties like magnetic susceptibility and specific heat, when the temperature is decreased. In contrast, the critical fluctuations are suppressed at low temperatures in the 1/1\,AC (Yb$_{14}$Au$_{51}$Al$_{35}$) \cite{deguchi2012quantum}. Thus, aperiodicity brings the Yb-Au-Al system closer to the quantum critical point. The question, however, remains whether quasicrystals are fundamentally different from their approximants in this regard.

The Yb-Au-Al intermetallic system contains a QC and a 1/1\,AC phase. Here, the focus is on the Ce-Au-Al system which hosts 2/1 and 1/1\,AC phases, whereas the QC phase is absent \cite{muro2021composition}. Muro et al.~\cite{muro2021composition} have reported on 1/1\,AC and 2/1\,AC specimens with various Au concentrations in the Ce-Au-Al system.
From their study of zero-field specific heat, magnetic susceptibility at 1\,T, and resistivity, they concluded that these systems show signatures of a Kondo resonance as well as spin-glass behavior below 1\,K.
Here, we study the effect of magnetic field on the ac-susceptibility and low-temperature specific heat of single crystal 1/1\,AC and 2/1\,AC samples. The ac-susceptibility measurements show a diverging behavior for low temperatures, and the magnetic field behavior closely resembles that of other quantum critical systems. The low-temperature specific heat is dominated by a magnetic Schottky anomaly due to a splitting of the Kramers doublet ground state of Ce$^{3+}$.

\section{Experimental Details} 
The samples by Muro et al.~\cite{muro2021composition} were synthesized using arc melting with stoichiometric compositions Ce\textsubscript{14}Au\textsubscript{x}Al\textsubscript{86-x }and Ce\textsubscript{16}Au\textsubscript{x}Al\textsubscript{84-x}, followed by annealing at 800\,°C for 50\,hours. Although homogeneous and polycrystalline samples were produced, an inseparable secondary phase, the hexagonal Ce\textsubscript{14}Au\textsubscript{51} compound 
\cite{Trovarelli1994}, coexisted in most of the samples. This secondary phase potentially contributed to the magnetic observations of the specimens. Therefore, to avoid the formation of the hexagonal Ce\textsubscript{14}Au\textsubscript{51 }compound, we applied a different synthesis approach aimed at isolating phase-pure, well-faceted single crystals with selected compositions of Ce\textsubscript{14}Au\textsubscript{71}Al\textsubscript{15} for 1/1\,AC and Ce\textsubscript{12}Au\textsubscript{65}Al\textsubscript{23} for 2/1\,AC.

\begin{figure*}[t!!]
\centering
\includegraphics[width=0.9\textwidth]{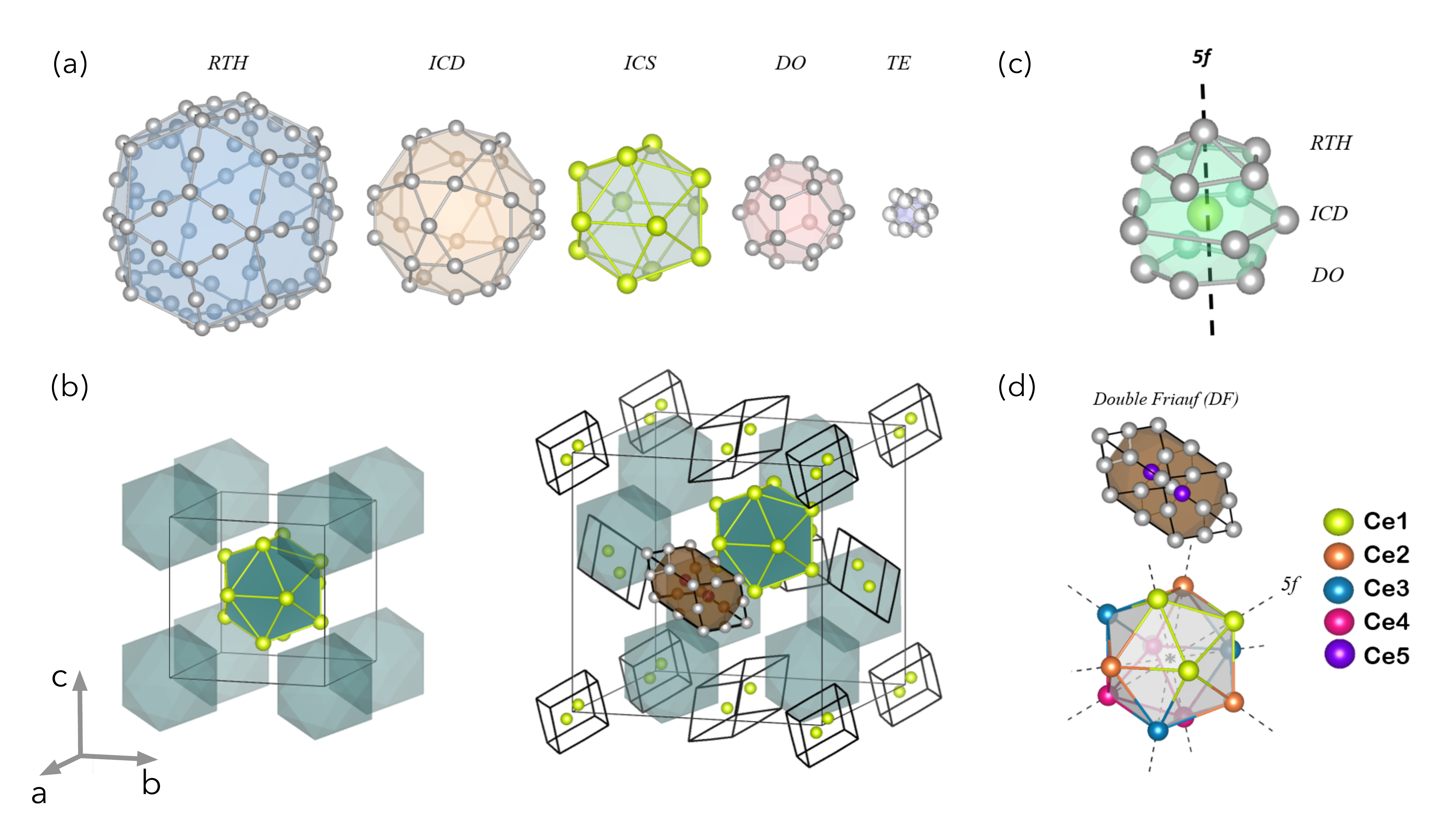} 
\caption{Motifs of the cubic 1/1 and 2/1\,AC structures in the Ce-Au-Al system. (a). Polyhedron shells of the Tsai cluster which is the fundamental unit of both structures. RTH: rhombic triacontahedron, ICD: icosidodecahedron, ICS: icosahedron, DO: dodecahedron, TE: tetrahedron. The ICS shell is exclusively built from Ce atoms, indicated in yellow. Grey circles signify pure Au, Al, or mixed Au/Al positions found throughout the various other shells. (b). Spatial distribution of ICS shells within the unit cells of the 1/1 and 2/1\,ACs.
In the 2/1 AC, which more closely relates to QC, an extra Ce atomic position is observed outside the ICS within the acute rhombohedron (AR) unit, marked by black lines, which fills the space between the RTH shells \cite{Gomez2001}. (c) and (d). Symmetry-independent Ce positions in the 2/1 AC, situated within both the ICS and the double Friauf (DF) polyhedral inside the AR. The various colored spheres in (d) represent distinct Ce atoms, with Ce1-4 in the ICS, and Ce5 in the DF. The ICS’s twelve vertices suggest twelve pseudo-5-fold rotational axes (5f) throughout the structure. Each vertex’s Ce atom is central within a 16-atom polyhedron, bridging the inner DO shell to the outer RTH shell along the pseudo 5f axis, as shown in (c).}
\label{fig:1}
\end{figure*}

Crystals of Ce$_{14}$Au$_{71}$Al$_{15}$ 1/1\,AC were precipitated from a melt with composition Ce$_{5}$Au$_{70}$Al$_{25}$ (about 2\,g total mass) which was initially heated to 1273\,K and subsequently cooled at a slow rate (2\,K/h) to 993\,K. Upon reaching this temperature crystals were separated from the melt by centrifugation \cite{Canfield2016}. Crystals of Ce$_{12}$Au$_{65}$Al$_{23}$ 2/1 AC were obtained from a mixture of Ce$_{10}$Au$_{66}$Al$_{24}$ (about 2\,g total mass) through the peritectic reaction of 1/1\,AC and melt at 973\,K upon annealing for 10\,days and subsequent isothermal centrifugation. The reaction mixture was initially heated to 1273\,K and then rapidly quenched in liquid nitrogen, following the procedure introduced in 
\cite{gebresenbut2021peritectic}. The starting materials were high-purity 99.99\,at.$\%$ granules of Ce, Au, and Al purchased from ChemPUR.

Structural characterization was performed using powder and single-crystal X-ray diffraction. For compositional analysis, energy-dispersive X-ray spectroscopy on cross-section polished samples was utilized using a Zeiss LEO 1550 scanning electron microscope (for details see Supplementary Fig.~S1 and Fig.~S2). The ac-susceptibility was measured using a  PPMS (Quantum Design Dynacool 9\,T ACMS II). Specific heat measurements were performed in a Bluefors dilution refrigerator using a membrane-based nanocalorimeter \cite{tagliati2012differential, willa2017nanocalorimeter} with a sample size of 20\,nmol for 1/1\,AC and 15\,nmol for 2/1\,AC. Measurements were performed in the ac-steady state mode with continuously adjusted frequency for good accuracy, differential thermometry running at the 5th harmonic of the temperature oscillation, and local temperature control on the calorimeter membrane with the dilution refrigerator sitting at base temperature.  

\section{Results and Discussion}

The structures of Ce-Au-Al ACs are built of so-called Tsai clusters which are the fundamental structural unit of icosahedral Tsai-type QCs \cite{tsai2000stable}.  
A Tsai cluster features a multi-shell arrangement of concentric polyhedra with the outermost shell being a rhombic triacontahedron (RTH), followed by an icosidodecahedron (ICD), an icosahedron (ICS), a dodecahedron (DO), and a tetrahedron (TE) at the center (Fig.~\ref{fig:1}a). In terms of crystallography, both 1/1 and 2/1\,ACs are periodic crystals in 3D space, each with a distinct cubic space group, which leads to different Tsai-cluster arrangements in their unit cells (Fig.~\ref{fig:1}b) \cite{Gomez2001, Takakura2007}. The 1/1\,AC typically forms a body-centered cubic (bcc) structure with the space group Im-3 (However, we noticed superstructure reflections for the Ce-Au-Al 1/1\,AC which suggest a lower symmetry, see Supplementary Fig.~S1). In contrast, the 2/1\,AC adopts a primitive cubic structure with the space group Pa-3. The 16-atom polyhedron, constructed from the vertices of the ICS along the pseudo-5f axis, is considered the coordination environment for the Ce atoms (Fig.~\ref{fig:1}c,d).

\begin{figure}[t!!]\includegraphics[width=0.9\columnwidth]{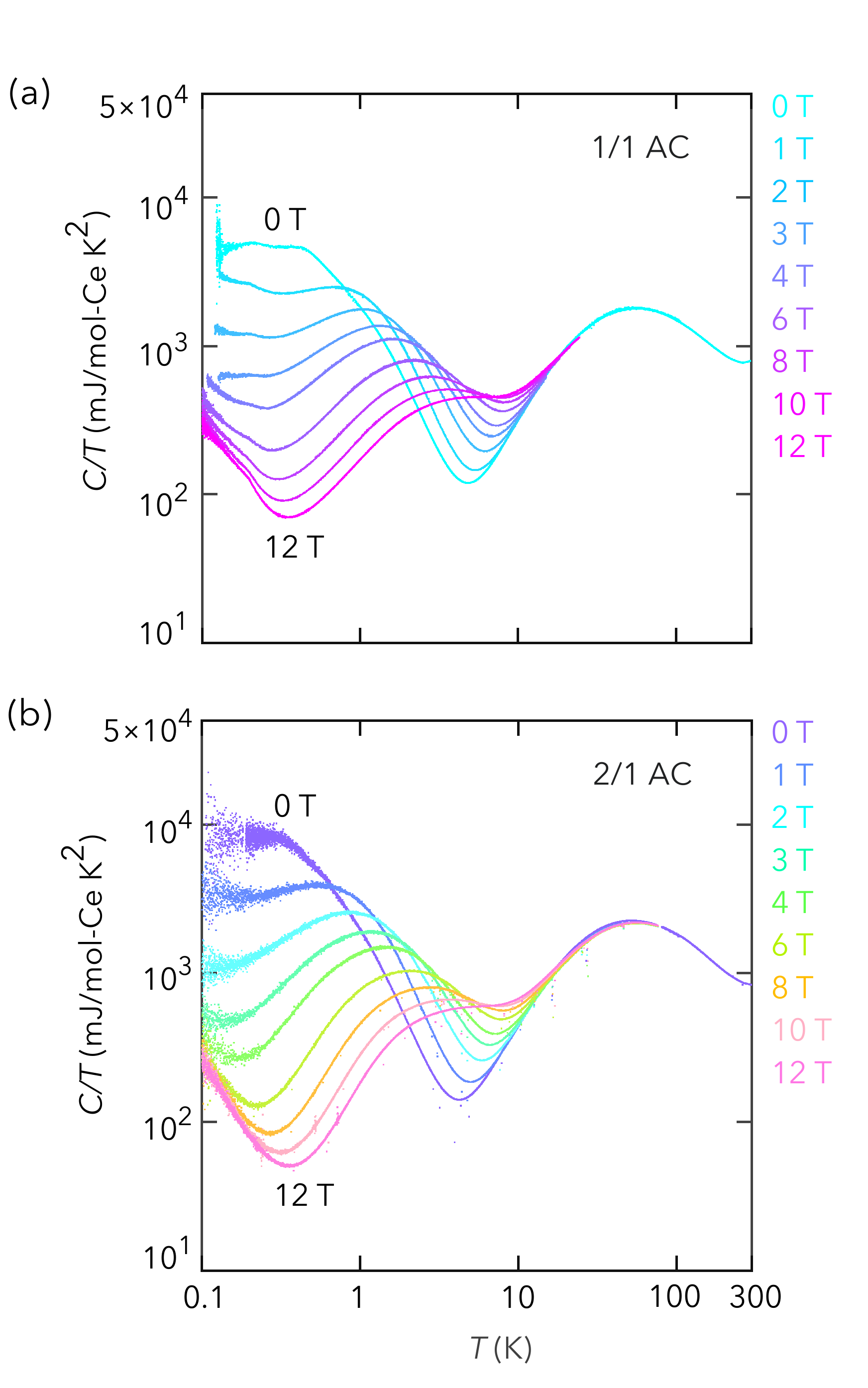} 
\caption{Specific heat of 1/1\,AC and 2/1\,AC in the Ce-Au-Al system. The specific heat is shown as $C/T$ per mol Ce for various magnetic fields for (a) 1/1 and (b) 2/1\,AC, respectively.}
\label{fig:2}
\end{figure}

Figure~\ref{fig:2} shows the specific heat of 1/1 and 2/1\,AC for several magnetic fields. The specific heats for 1/1 and 2/1\,AC are quite similar.
At high temperatures, above $\sim 10$\,K the behavior is dominated by the phonons. At the lowest temperatures, below 0.3\,K, the upturn in $C/T$ is due to a nuclear specific heat arising from $^{197}$Au with nuclear spin $I=3/2$ and $^{27}$Al with $I=5/2$ \cite{Stone2016}, apparent at high magnetic fields. The dominant contribution of the low-temperature specific heat resembles a magnetic Schottky anomaly, shifting up in temperature with increasing magnetic field.

\begin{figure}[t!!]\includegraphics[width=1.0\columnwidth]{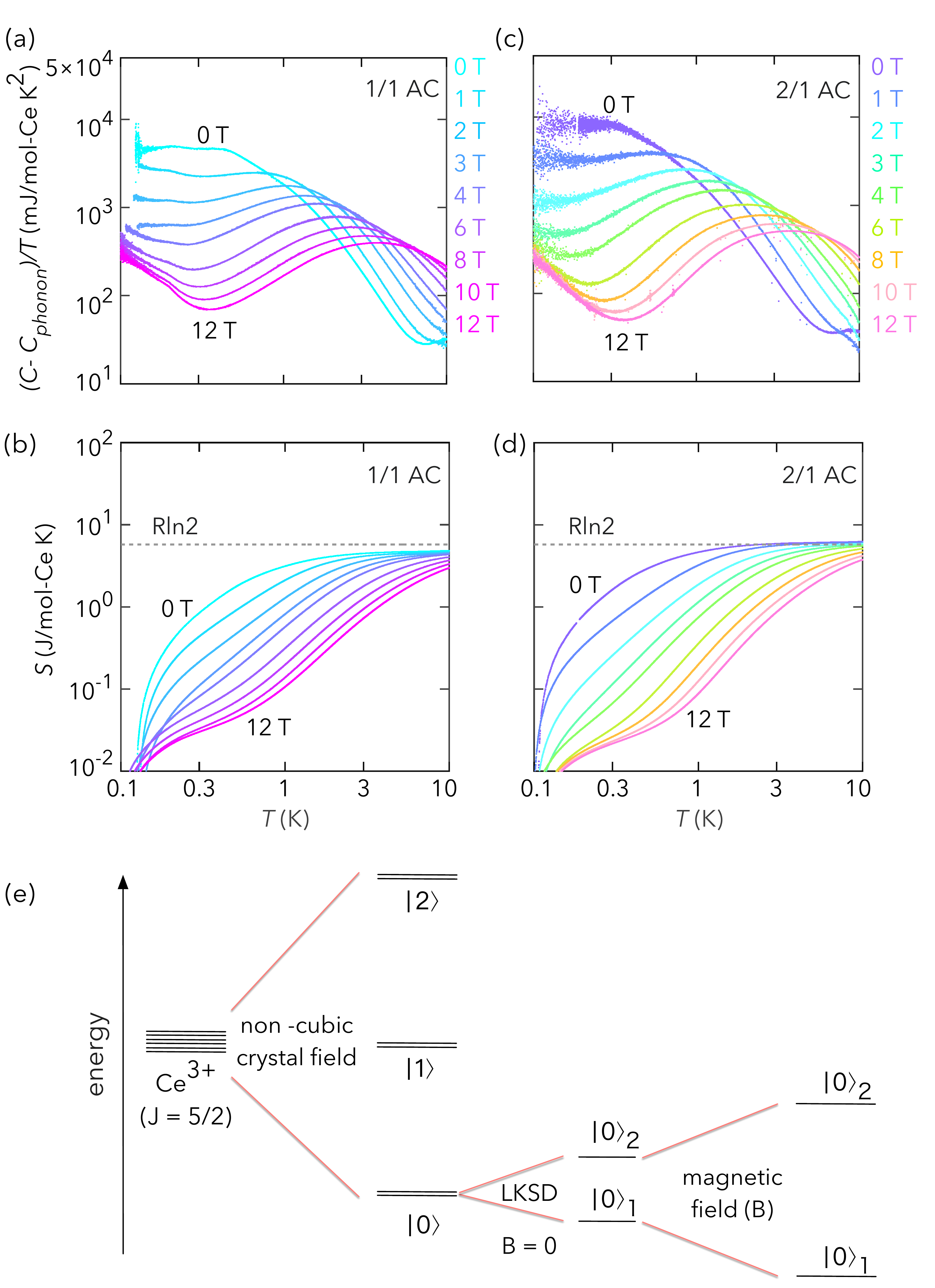} 
\caption{Specific heat and entropy of 1/1\,AC and 2/1\,AC. 
(a) and (c). The specific heat after removing the phonon contribution (see Supplementary Fig.~S3 and S4) is shown as $C/T$ for 1/1 and 2/1\,AC, respectively. 
(b) and (d). Corresponding entropy of the systems. The dotted gray line indicates the value $R$ln2. 
(e). Sketch of splitting of Ce$^{3+}$ ($J$=5/2) degenerate levels, due to the non-cubic crystalline field, into three Kramers doublets. Crystal field disorder or magnetic interactions in the system could be the reason for lifting the Kramers spin degeneracy (LKSD). The magnetic field further splits the ground state doublet $|0\rangle$ as seen in the panel e.}
\label{fig:3}
\end{figure}

Figure~\ref{fig:3}a and c show the specific heat for the 1/1 and 2/1\,ACs after subtracting the phonon part, approximated using the Debye $T^3$ law. 
\begin{figure*}[t!!]\includegraphics[width=0.95\linewidth]{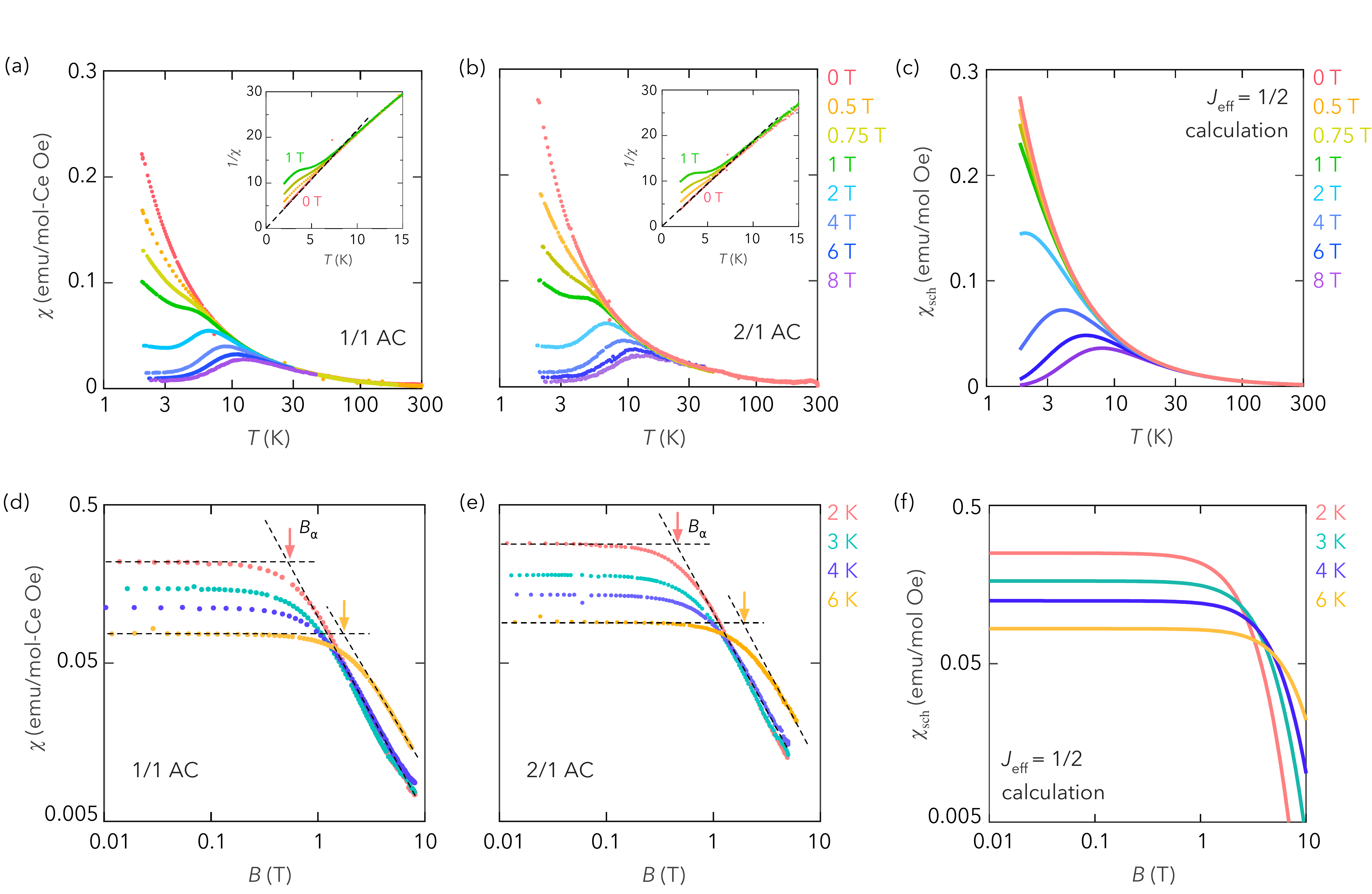} 
\caption{Temperature and magnetic field dependence of the ac susceptibility.
(a) and (b). The ac-susceptibility ($\chi$) for 1/1 and 2/1\,AC for various magnetic fields in the log-lin scale. The inset shows the inverse susceptibility for several magnetic fields at low temperatures. The dotted line corresponds to $1/\chi \propto T$.
(c). Calculated temperature dependence of the susceptibility for a two-level system for magnetic fields corresponding to panel a and b.
(d) and (e). Magnetic field dependence of ac-susceptibility ($\chi$) for several temperatures in log-log scale. The applied magnetic field cuts off this diverging behavior and $\chi$ starts to decrease above the cut-off field ($B_{\alpha}$), shown as arrows. 
(f). Corresponding calculation for the field dependence of susceptibility for a two-level system for magnetic fields in panel d and e.
}
\label{fig:4}
\end{figure*}
As we apply the magnetic field, the peak in $C/T$ shifts to higher temperature. This magnetic field behavior of the temperature dependence shows a maximum in $C$ that takes approximately the same value for all applied magnetic fields (see Supplementary Fig.~S4). This indicates that the behavior is originating from a Schottky anomaly.
The corresponding entropies, obtained as the integral of $C/T$, are shown in Fig.~\ref{fig:3}b and d. As seen from the entropy analysis, the entropy saturates to a value close to $\sim R\ln 2$ for both ACs at around 10\,K. This indicates that the Schottky anomaly arises from a two-level system.

The non-cubic crystal field splits the Ce$^{3+}$ ($J=5/2$) ground state into three Kramers doublets \cite{kramers1930, Wigner1993, Desgranges2014, Griffith2019}. A sketch of the energy level splitting is shown in Fig.~\ref{fig:3}e, where the ground state Kramers doublet is denoted as $|0\rangle$. The external magnetic field further splits the ground state doublet due to the Zeeman effect. This effect of the magnetic field on the Kramers doublet is seen in the specific heat in Fig.~\ref{fig:3}a and c. This ground state Kramers doublet, thus, determines the low-temperature behavior in specific heat.

From the crystal structure studies, it is seen that Ce has five distinct crystal environments (Fig.~\ref{fig:1}d). These environments may induce different crystal-field splittings for Ce$^{3+}$, but will all have a Kramers doublet ground state. The Zeeman splitting of the Kramers doublet in magnetic field will be different for the different atomic sites, since the magnetic field will have different relative orientations with respect to the principal magnetic axes of the doublets. This makes it difficult to model details of the temperature and field dependence of the specific heat. Compared to a single two-level system with an energy gap proportional to the magnetic field, the current AC systems display a broader peak in specific heat.

The presence of a significant zero-field specific heat indicates that the Kramers doublet ground state is not degenerate in the absence of magnetic field. This lifting of Kramers degeneracy could result from crystal field disorder \cite{krempasky2024altermagnetic} or magnetic interactions within the system \cite{Winkler2003, Vsmejkal2022, Vsmejkal2022_2nd, Lee2024, krempasky2024altermagnetic}.

Figure~\ref{fig:4}a and b show the temperature dependence of the ac-susceptibility $\chi$ of 1/1 and 2/1\,AC, respectively, for magnetic fields up to 8\,T.
The high-temperature (10-150\,K) ac-susceptibility can be explained by the Curie-Weiss law ${\chi = W/(T-\theta)}$
where $\theta$ is the Curie temperature and 
\begin{align}\label{Eq: Curie_constant}
    W = \frac{\mu_0 N_{\mathrm{A}} \mu^2}{3k_{\mathrm{B}}}
\end{align}
is the Curie constant, and $\mu = g\sqrt{J(J+1)}\mu_\mathrm{B}$ where $g$ is the Land\'e g-factor.
Note that in this temperature range $\chi \sim 0.005$\,emu/mol-Ce\,Oe (see Supplementary Fig.~S5) which is significantly larger than any background from diamagnetic or electronic contributions \cite{Mugiraneza2022, Van1965}.
The Curie temperature extracted from a fit of $1/\chi$ vs $T$, shown in Supplementary Fig.~S5, is $-10\pm0.5\,\mathrm{K}$ and $-6\pm1\,\mathrm{K}$ for 1/1 and 2/1\,ACs, respectively, indicating weakly antiferromagnetic coupling. The effective magnetic moment $\mu_{\mathrm{exp}}$ of Ce obtained from this high-temperature region is  $\mu_{\mathrm{exp}} = 2.53\pm0.02\,\mu_{\mathrm{B}}$ and $\mu_{\mathrm{exp}} = 2.50\pm0.02\,\mu_{\mathrm{B}}$ for 1/1 and 2/1\,ACs, respectively (see Supplementary Fig.~S5). This effective magnetic moment is close to the moment for free Ce$^{3+}$ ion ($\mu_\mathrm{cal} = 2.54\,\mu_{\mathrm{B}}$).

Going to lower temperatures, the zero-field susceptibility no longer follows the high-temperature Curie-Weiss law. The insets of Fig.~\ref{fig:4}a and \ref{fig:4}b show the low-temperature inverse susceptibility.
A Curie-Weiss fit for temperatures between 2-10\,K yields $\mu_{\mathrm{exp}} = 1.98\pm0.02\,\mu_{\mathrm{B}}$ and $\mu_{\mathrm{exp}} = 2.10\pm0.02\,\mu_{\mathrm{B}}$ for 1/1 and 2/1\,ACs, respectively (see Supplementary Fig.~S5), with Curie temperatures well below 1\,K. The low-temperature zero-field $\chi$, thus, shows essentially a power-law behavior ${\chi \propto 1/T}$.
The fairly large $\mu_{\mathrm{exp}}$ implies that the ground state Kramers doublet has a significant contribution of $m_{J}= 5/2$.

The diverging behavior in $\chi(T)$ is suppressed when an external magnetic field is applied, as seen in Fig.~\ref{fig:4}a and b. Figure~\ref{fig:4}d and \ref{fig:4}e show the magnetic field dependence of 1/1 and 2/1\,AC, respectively, for several temperatures. For a given temperature there exists a cutoff field $B_{\alpha}$ above which this {\bluetext{suppression}} happens. Well below $B_{\alpha}$, the susceptibility becomes independent of the magnetic field and follows the zero-field $\chi(T)$. 
Above $B_{\alpha}$, the decrease of $\chi$ with field follows a power-law behavior. At the highest fields and lowest temperatures, the behavior becomes independent of temperature{\redtext{ and the magnetic field controls the cutoff behavior}}{\bluetext{, while at low fields the behavior becomes independent of field. This indicates a temperature-field competition that governs the behavior of $\chi$}}.

The diverging zero-field susceptibility $\chi \propto 1/T$ could naively be attributed to a simple, non-interacting paramagnet with $\theta = 0$.
From the entropy in Fig.~\ref{fig:3}b and ~\ref{fig:3}d we can see that the low temperature behavior of the system comes from the ground state Kramers doublet. The {\redtext{ground state Kramers}} doublet acts as an effective spin 1/2 system, thus, $J_\mathrm{eff} = 1/2$ at low temperatures \cite{Alonso2015, Bordelon2019}.
The ac-susceptibility of a $J_\mathrm{eff} = 1/2$ Kramers doublet system in magnetic field is given by
\begin{align}\label{Eq:Schottky_Chi}
    \chi_\mathrm{sch} = \frac{W_\mathrm{K}}{T}& \left[1- \tanh^2\left({\frac{g_\mathrm{K}\mu_\mathrm{B}B}{2k_\mathrm{B}T}}\right) \right] 
\end{align}
where 
\begin{align}\label{Eq:Schottky_Chi_W}
    W_\mathrm{K} = \frac{\mu_0 N g_\mathrm{K}^2 \mu_\mathrm{B}^2}{4k_\mathrm{B}}. 
\end{align}
Note that the $g$-factor in Eq.~(\ref{Eq:Schottky_Chi}) corresponds to the Kramers doublet and not the Land\'e $g$-factor. The zero field low-temperature fit of $1/\chi$ vs.\ $T$ (see Supplementary Fig.~S5) directly gives $W_\mathrm{K} \approx 0.5$, which in turn gives $g_\mathrm{K} \approx 2.3$. The temperature and field dependence of the susceptibility for the ground state Kramers doublet from Eq.~(\ref{Eq:Schottky_Chi}) using $g_\mathrm{K} = 2.3$ and $W_\mathrm{K} = 0.5$\,(emu\,K/Oe)/mol is shown in Fig.~\ref{fig:4}c and \ref{fig:4}d for the corresponding fields and temperatures.

For a quantum critical system, the susceptibility has a form of ${1/\chi \propto T^\beta}$, where $\beta$ is a critical exponent. This critical exponent characterizes the spectrum of critical fluctuations in a quantum critical system \cite{lohneysen2007fermi}. For Ce-Au-Al ACs at low temperatures (2-10\,K), this critical exponent seems to be approximately equal to one ($\beta = 0.94 \pm 0.01$ for 1/1 AC and $0.97 \pm 0.01$ for 2/1 AC).
This suggests that $\beta \approx 1$ and most of the behavior seen in ac-susceptibility comes from the Zeeman splitting of the Kramers doublet in magnetic fields similar to the behavior seen in specific heat.

From transport measurements \cite{muro2021composition}, it is seen that the ACs of the Ce-Au-Al show a Kondo resonance peak at $T_\mathrm{K}\approx $1\,K.
This indicates that the interaction between the conduction electrons and $f$-electrons becomes important only below 1\,K. The conduction electron effective mass above 5\,K as extracted from the Sommerfeld coefficient evaluated in Supplementary Fig.~S3 is rather modest, with $\gamma \approx 4$\,mJ/mol-at.K$^2$. Thus, the system does not display any significant heavy fermion characteristics, consistent with $T_\mathrm{K}$ being small.

{\redtext{From the temperature and field dependence of low-temperature $\chi$, it can be seen that $\chi$ shows a temperature-field competition that sets the cutoff scale in this region. This}}{\bluetext{The temperature-field competition}} is one of the defining features of quantum criticality in a broad range of systems \cite{Millis2002, lohneysen2007fermi, Gegenwart2008, Matsumoto2011, Wu2018, Khansili2023} including the quantum critical quasicrystal Yb-Au-Al \cite{deguchi2012quantum, Watanuki2012, Shaginyan2013}. 
However, the susceptibility of even a non-interacting two-level system behaves very similarly in magnetic fields, as seen from Figs.~\ref{fig:4}c and \ref{fig:4}f. There are differences, though. The measured $\chi$ approximately follows a power-law behavior in high fields, which is an indication of quantum critical behavior, whereas the $\chi_\mathrm{sch}$ calculations appear as a more sudden drop above the cutoff field.

Most of the heavy fermion systems have a Kramers doublet ground state \cite{Coleman2007, Coleman2010}. The strong interaction of the conduction electrons with the $f$-electrons induces the quantum critical behavior observed in these systems \cite{Coleman2007, Coleman2010}. In heavy fermion quantum critical systems, the critical fluctuations modify this behavior of the Kramers doublet by changing the $\beta$ and significantly reducing the $g_\mathrm{K}$ of the doublet. {\bluetext{Indications of interactions in the Ce-Au-Al approximants studied here are seen both at zero field at at high field. As seen from $C/T$, the Kramers degeneracy is lifted even in zero magnetic field. At high fields, the power-law dependence of $\chi$ gives a weaker field dependence than expected from a non-interacting system.
}}

From the present data, it is not clear whether the ground state of the system is magnetically ordered or a paramagnetic state. If magnetic ordering sets in, the transition temperature is below about 0.2\,K, since no clear signatures are seen in specific heat.
The reason for this could be the geometric frustration from the ACs structure. {\bluetext{The complex atomic clusters allow multiple rare-earth sites with different crystal fields and orientation with respect to the applied field. This likely contributes to the suppression of long-range magnetic order in the system.}}

\section{Conclusions}
We have studied the effect of magnetic field on the low-temperature magnetic susceptibility and specific heat capacity for the 1/1 and 2/1 ACs of the Ce-Au-Al system. The systems display diverging magnetic susceptibility at zero fields.
Even though the measured ac-susceptibility in magnetic field closely resembles the behavior of a quantum critical system, the major characteristics can be explained by the splitting of the ground state Kramers doublet in magnetic field. The reason for this resemblance is that the Kramers doublet acts as a precursor for the heavy fermion quantum criticality.
Specific heat does not find any magnetic ordering, but also does not show any clear signatures of critical behavior.
Further investigations are needed to rule out magnetic order below 0.2\,K. We anticipate that stronger interaction of conduction electrons with the Kramers doublet will lead to modification of {\bluetext{both the exponent $\beta$ and the $g$-factor}} $g_\mathrm{K}$ of the doublet such that it develops a strong temperature dependence. The corresponding redistribution of entropy will be an indication of critical fluctuations and a quantum critical ground state. {\bluetext{By substituting Ce with Yb, stronger interactions are induced, and a quantum critical state is achieved, both in the ACs and QCs \cite{Watanuki2012, deguchi2012quantum}. Partial substitution of this sort is anticipated to illustrate how the current precursor state becomes quantum critical.}}

\section{Acknowledgments}
\begin{acknowledgments}
This work was supported by the Knut and Alice Wallenberg Foundation (Grant KAW 2018.0019). A.K. and A.R. acknowledge support from the Swedish Research Council, D.\,Nr.\,2021-04360.
\end{acknowledgments}

\bibliographystyle{apsrev4-1}
\bibliography{Khansili_CeAuAl.bib}



\renewcommand{\thefigure}{S\arabic{figure}}
\setcounter{figure}{0}

\section{Supplementary Figures}

\begin{figure*}[h!]
\includegraphics[width=0.9\textwidth]{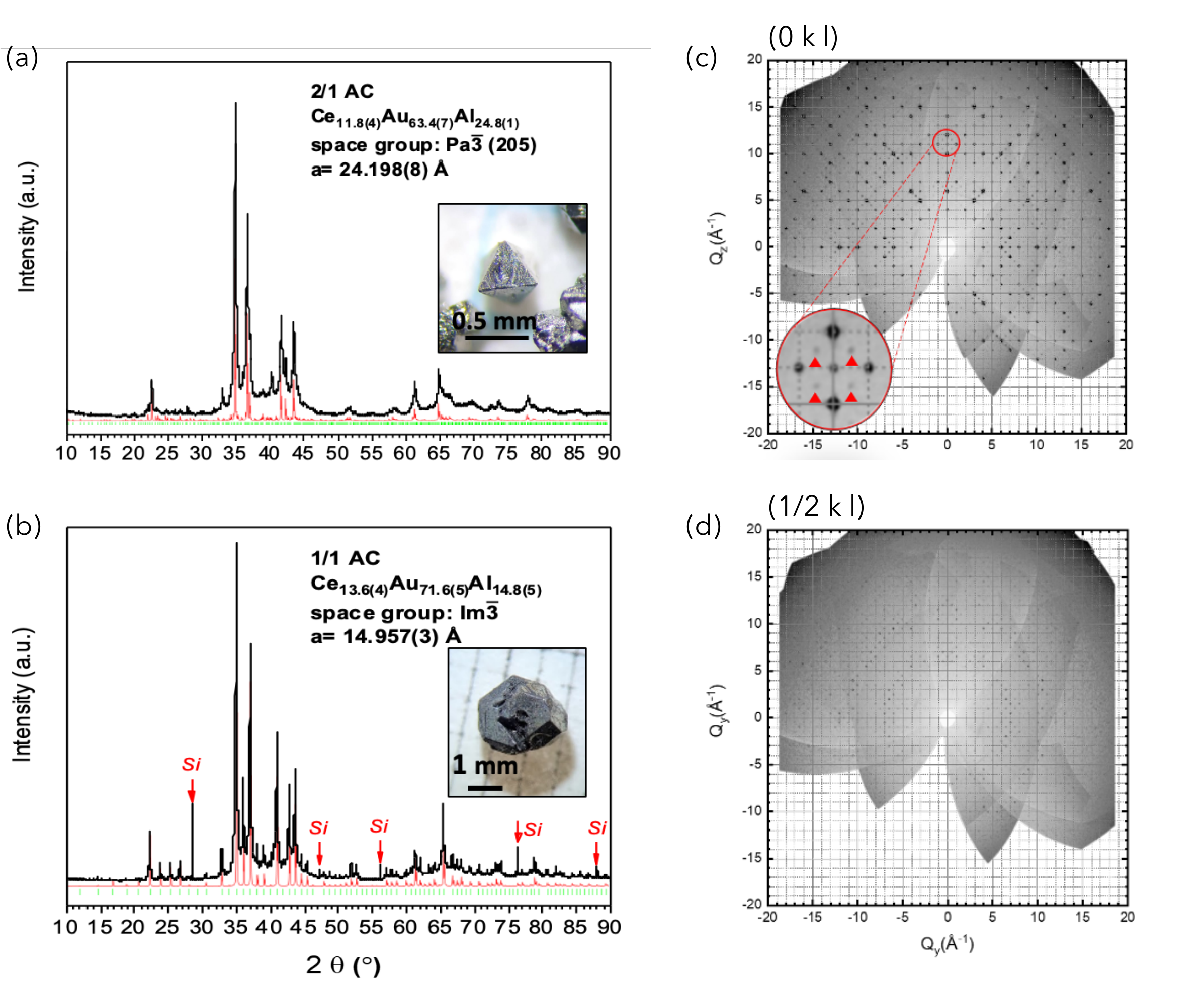} 
\caption{(a). Comparison of the experimental PXRD pattern (black curve) and calculated PXRD pattern (red curve) of Ce\textsubscript{12}Au\textsubscript{65}Al\textsubscript{23} 2/1\,AC. (b). Comparison of the experimental PXRD pattern (black curve) and calculated PXRD pattern of the basis structure of Ce\textsubscript{14}Au\textsubscript{71}Al\textsubscript{15} 1/1\,AC. The insets in (a) and (b) report the  EDS determined compositions and refined lattice parameters, and show photographs of well-faceted, mm-sized sample specimens obtained from the applied synthesis procedure. Reflections from Si standard are marked with red arrows.
(c) and (d). SCXRD data of 1/1\,AC crystals show superlattice reflections (red triangles in panel c) in both the 0kl and 1/2 kl sections, respectively. The grid size is with respect to the basic 1/1\,AC unit cell, a $\approx$ 14.9\,\AA. To index all the reflections the unit cell of the basis structure has to be doubled in all directions a,b,c.} 
\label{fig:SI1}
\end{figure*}

\begin{figure*}[ht]
\includegraphics[width=0.9\textwidth]{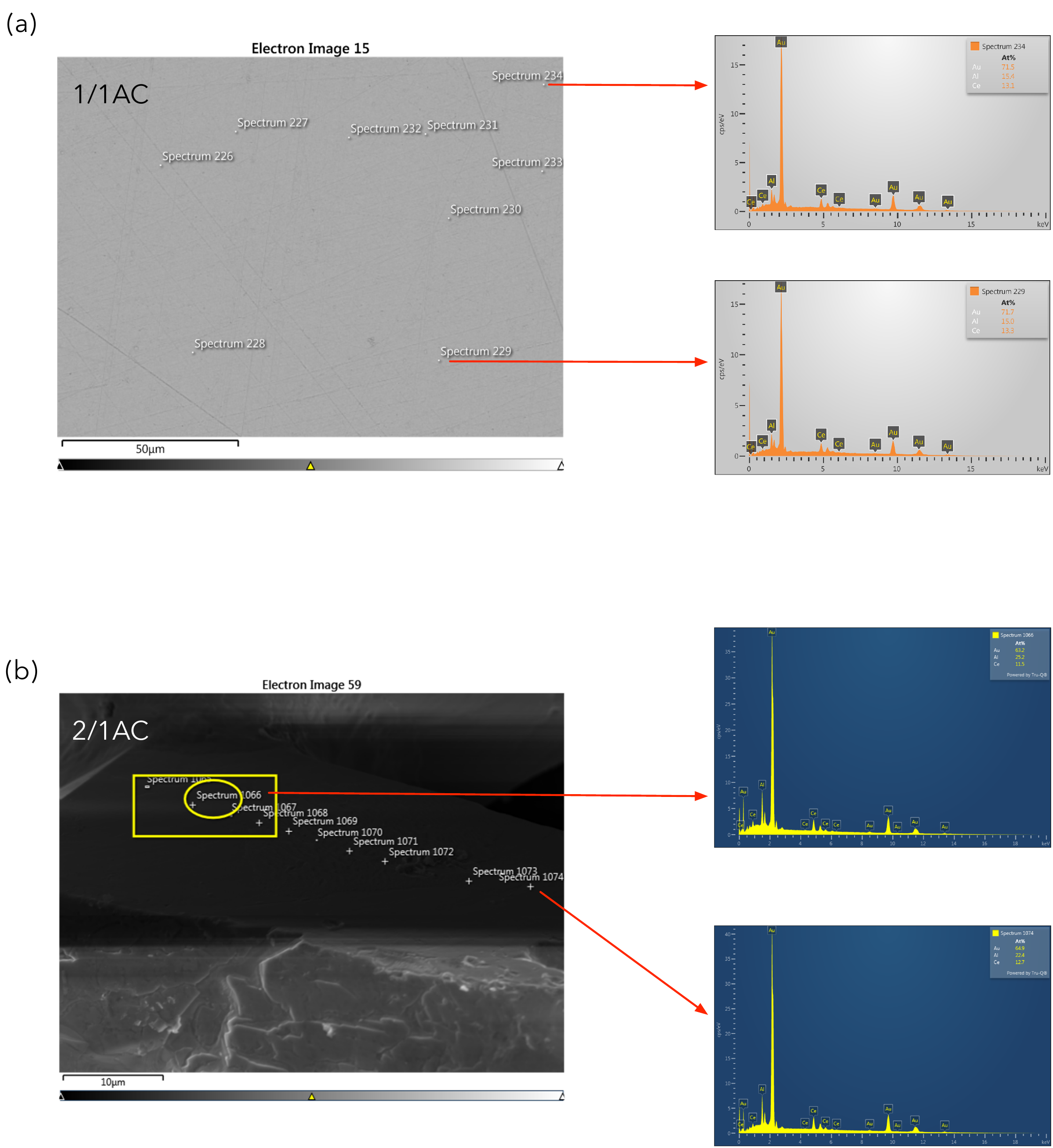} 
\caption{EDS spectra for 1/1AC (a) and 2/1AC (b).}
\label{fig:SI_EDS}
\end{figure*}

\begin{figure*}[ht]
\includegraphics[width=0.98\textwidth]{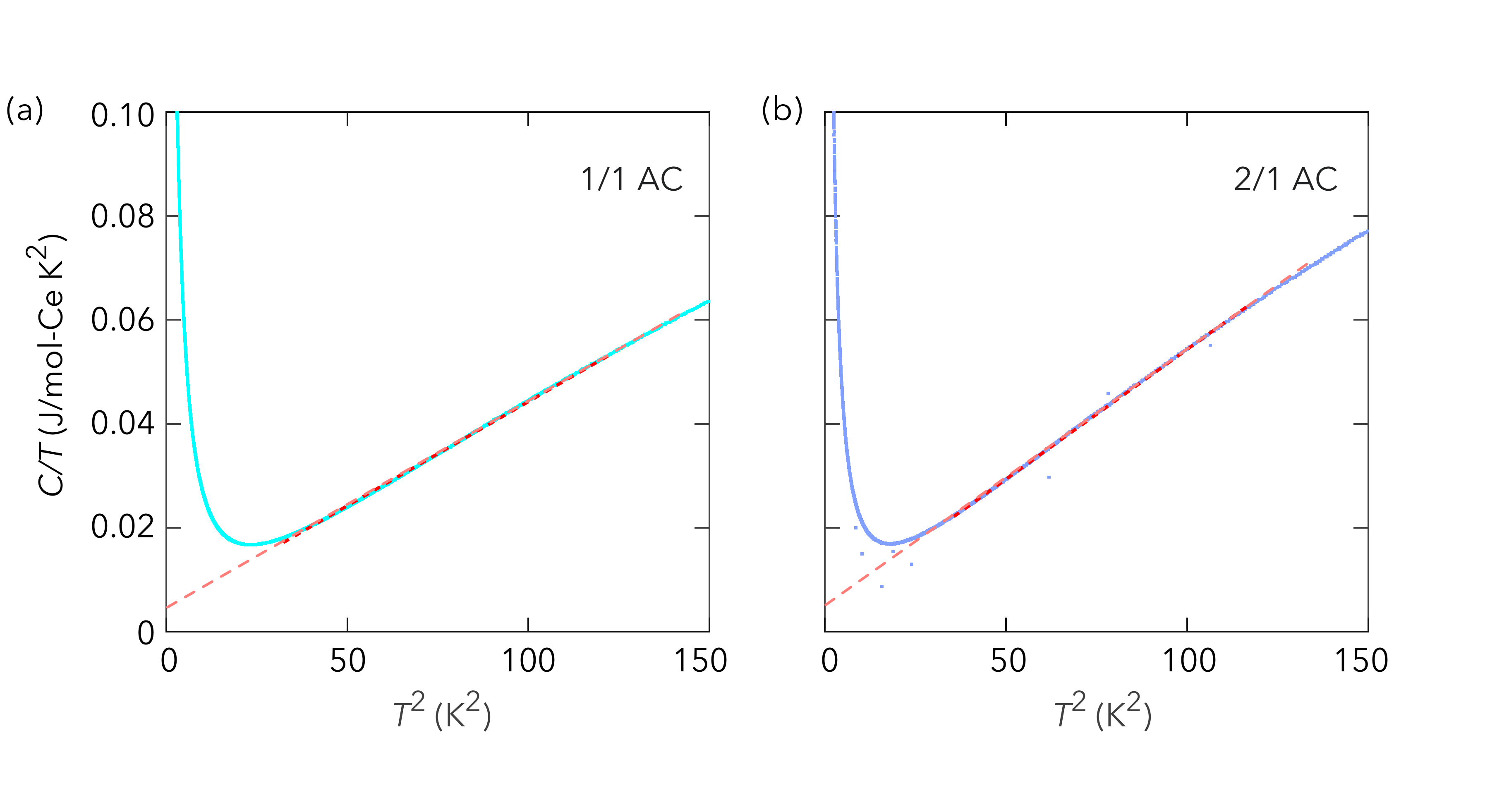} 
\caption{Specific heat as $C/T$ vs. $T^2$. The red dashed line represents a linear fit that gives the conduction electron and phonon specific heat $(C/T) = \gamma + \beta T^2$. The extrapolated Sommerfeld coefficient $\gamma$ is about 4\,mJ/mol-atomK$^2$ for both compounds, and the Debye coefficient $\beta$ is 0.40 and 0.50\,mJ/mol-atom\,K$^4$ for 1/1\,AC and 2/1\,AC, respectively, corresponding to Debye temperatures of 170\,K and 160\,K.}
\label{fig:SI2}
\end{figure*}

\begin{figure*}[ht]
\includegraphics[width=0.98\textwidth]{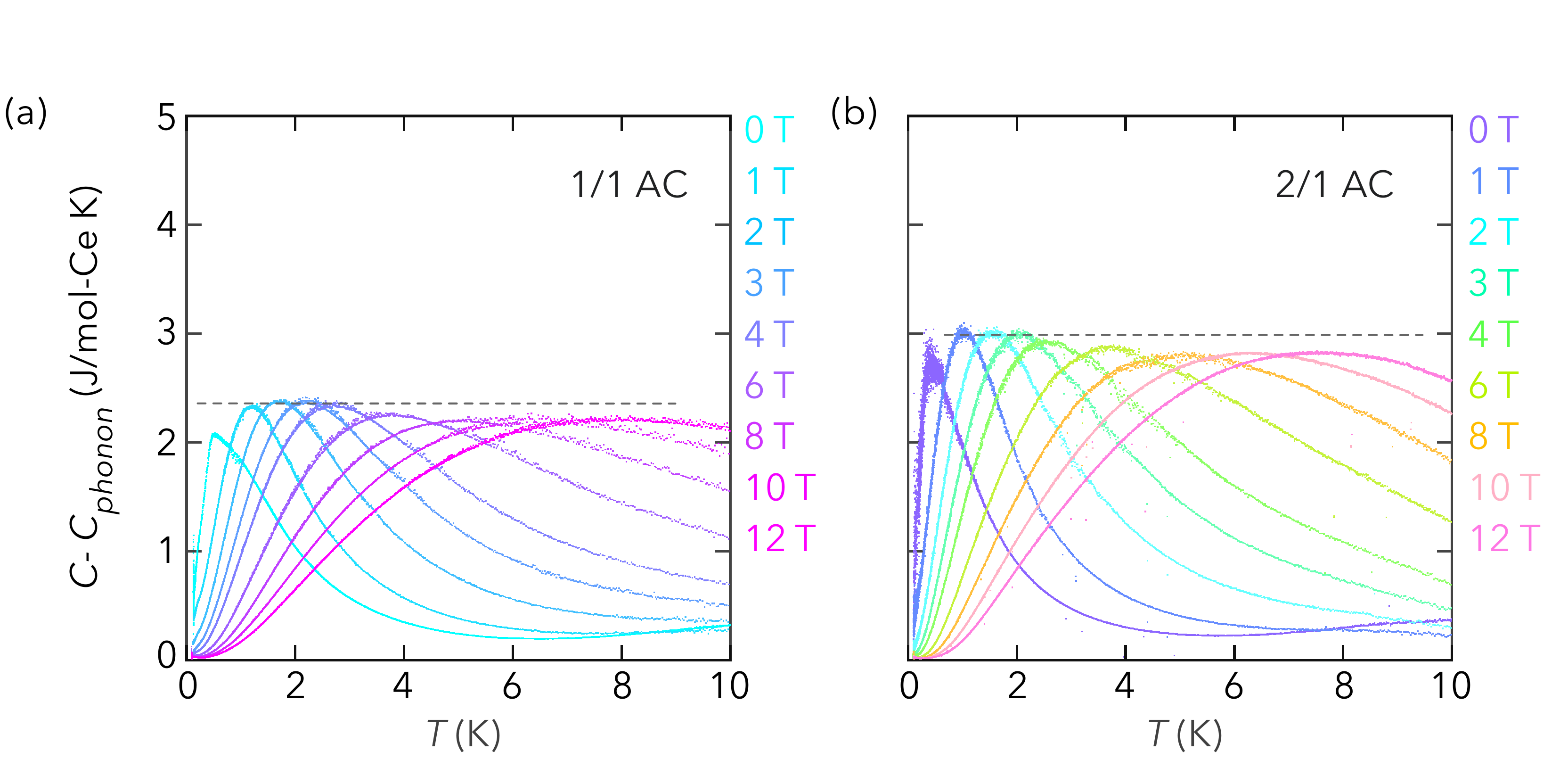} 
\caption{Schottky specific heat after removing the phonon contribution obtained from Fig.~\ref{fig:SI2}.} 
\label{fig:SI3}
\end{figure*}

\begin{figure*}[ht]
\includegraphics[width=0.9\textwidth]{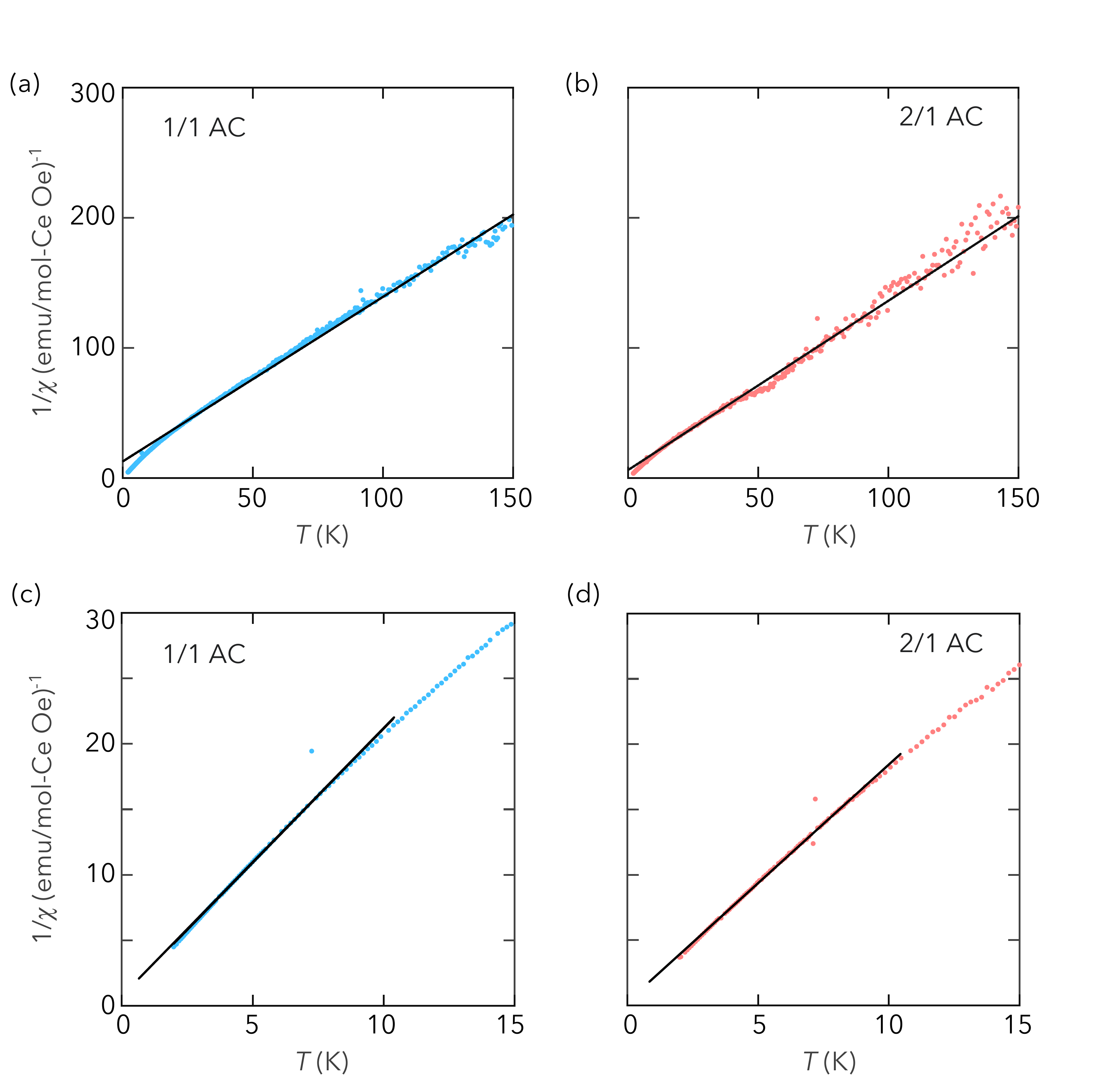} 
\caption{Zero-field inverse ac-susceptibility of 1/1AC and 2/1\,AC. (a) and (b). Solid line represent Curie-Weiss fit for the temperature range $10$-$150$\,K. The data is fitted to $1/\chi = (T - \theta)/W$. Here, ${W = {\mu_0 N_{\mathrm{A}} g_{\mathrm{eff}}^2 \mu_{\mathrm{B}}^2}/{3k_{\mathrm{B}}}}$ The fitting parameters are found to be $\theta = -10\pm0.5$\,K with $W= 0.80\pm0.01$\,(emu\,K/Oe)/mol-Ce for the 1/1 AC and $\theta =-6\pm1$\,K with $W= 0.78\pm0.01$\,(emu\,K/Oe)/mol-Ce for the 2/1\,AC.
(c) and (d). Curie-Weiss fit for the temperature range 2-10\,K. Fit gives $\theta = -0.32\pm0.06$\,K with $W= 0.487\pm0.005$\,(emu\,K/Oe)/mol-Ce for the 1/1 AC, and  $\theta = +0.14\pm0.05$\,K with $W= 0.549\pm0.005$\,(emu\,K/Oe)/mol-Ce for the 2/1\,AC.} 
\label{fig:SI4}
\end{figure*}

\end{document}